\documentclass{llncs}
\usepackage{graphicx,amsmath,amssymb,bbm,mathtools}

\usepackage{url}
\usepackage{breakurl}

\newcommand{\B}{{{\mathcal{A}}^{\star}}}
\newcommand{\G}{{\mathcal A}}
\newcommand{\f}{{\mathbf f}}

\begin{document}

\mainmatter              
\title{An example of degenerate hyperbolicity in\\ a cellular automaton with 3 states}
\titlerunning{Degenerate hyperbolicity in CA}  
%
\author{Henryk Fuk\'s \and Joel Midgley-Volpato}
\authorrunning{Henryk Fuk\'s et al.} 
\institute{Department of Mathematics and Statistics, Brock University\\
St. Catharines, Ontario L2S 3A1, Canada 
}

\maketitle              

\begin{abstract}
We show that a behaviour analogous to degenerate hyperbolicity can occur in nearest-neighbour
cellular automata (CA) with three states. We construct a 3-state rule by ``lifting'' elementary
CA rule 140. Such ``lifted'' rule is equivalent to rule 140 when arguments are
restricted to two symbols, otherwise it behaves as identity. We analyze the structure of
multi-step preimages of 0, 1 and 2 under this rule by using minimal finite state machines (FSM), and exploit
regularities found in these FSM. This allows to construct explicit expressions for densities of 
0s and 1s after $n$ iterations of the rule starting from Bernoulli distribution.
When the initial Bernoulli distribution is symmetric, the densities of all three symbols
converge to their stationary values in linearly-exponential fashion, similarly as
in finite-dimensional dynamical systems with hyperbolic fixed point with degenerate eigenvalues.
\end{abstract}
\section{Introduction}
In a linear continuous-time dynamical system given by $\dot{\mathbf{x}}=A\mathbf{x}$, if 
 $\mathbf{x}:\mathbb{R} \to
\mathbb{R}^n$ and $A$ is a real $n \times n$ matrix with all eigenvalues distinct and having negative real parts, $\mathbf{x}(t)$ tends to zero
exponentially fast as $t \to \infty$. The same phenomenon can be observed in nonlinear system
$\dot{\mathbf{x}}=\mathbf{f}(\mathbf{x})$ (where $\mathbf{f}:\mathbb{R}^n\to\mathbb{R}^n$)  in a vicinity
of hyperbolic fixed point, as long as the Jacobian matrix of $\mathbf{f}$ evaluated at the fixed point has only
distinct eigenvalues with negative real parts. If, on the other hand, the matrix $A$ has degenerate (repeated) eigenvalues, the convergence to the fixed point can be polynomial-exponential, that is, of the form $P(t)e^{-bt}$, where
$P(t)$ is a polynomial and $b>0$.

In discrete-time dynamical systems things are quite similar. For example, the linear system system 
\begin{equation}
\left[\begin{array}{c}
 x_{n+1}\\
 y_{n+1}
\end{array} \right]=
\left[\begin{array}{rc}
 0 & 1\\
 -\frac{1}{4} & 1
\end{array} \right]
\left[\begin{array}{c}
 x_{n}\\
 y_{n}
\end{array} \right]
\end{equation}
is defined by a matrix which has degenerate (double) eigenvalue $\frac{1}{2}$, thus polynomial-exponential (linear-exponential
in this case) convergence to the fixed point $(0,0)$ is expected. Indeed, the solution is
\begin{equation}
\left[\begin{array}{c}
 x_{n}\\
 y_{n}
\end{array} \right]=\left(\frac{1}{2} \right)^n
\left[\begin{array}{cc}
 1-n & 2n\\
 -\frac{n}{2}& 1+n
\end{array} \right]
\left[\begin{array}{c}
 x_{0}\\
 y_{0}
\end{array} \right],
\end{equation}
and we can clearly see the aforementioned linear-exponential convergence.

Cellular automata are infinitely-dimensional dynamical systems, yet a behaviour similar to hyperbolicity
in finite-dimensional systems has been observed in many of them. In particular, in some
binary  cellular automata in one dimension, known as \emph{asymptotic emulators of identity},   if the initial configuration is drawn from a Bernoulli distribution, 
the expected proportion of ones (or zeros) tends to its stationary value
exponentially fast \cite{paper52}.

Furthermore, an example of a probabilistic CA has been recently found \cite{probcabookchapter} where the density of ones converges to 
its stationary value in a linear-exponential fashion, just like in the case of degenerate hyperbolic fixed points
in finite-dimensional dynamical systems. Could such behaviour be observed in \emph{deterministic} CA as well? The purpose of this paper is to provide an example of such deterministic CA. 

We will consider 3-state nearest-neighbour CA obtained from elementary binary CA by ``lifting'' them to 
3-states.  What we mean by this is the following construction. Let $g:\{0,1\}^3 \to \{0,1\}$ be a
local function of elementary CA satisfying $g(0,0,0)=0$, $g(1,1,1)=1$, and let $f_g:\{0,1,2\}^3 \to \{0,1,2\}$ be defined by
\begin{equation} \label{defoflift}
 f_g(x_1,x_2,x_3)=
\begin{cases}
g(x_1,x_2,x_3), &  x_1,x_2,x_3 \in \{0,1\} \\
2g\left(\frac{x_1}{2},\frac{x_2}{2},\frac{x_3}{2}\right), &  x_1,x_2,x_3 \in \{0,2\} \\
g(x_1-1,x_2-1,x_3-1)+1, &  x_1,x_2,x_3 \in \{1,2\}\\
x_2, &\mathrm{otherwise}.
\end{cases}
\end{equation}
This construction ensures that when $f_g$ is restricted to two symbols only, it becomes equivalent to
$g$, otherwise it behaves as identity. Conditions $g(0,0,0)=0$, $g(1,1,1)=1$ ensure that there are no conflicts, so that, for example,
$f(2,2,2)$ is the same no matter if we apply second or third case of eq. (\ref{defoflift}).

\begin{figure}[t]
\begin{center}
\includegraphics[scale=0.7]{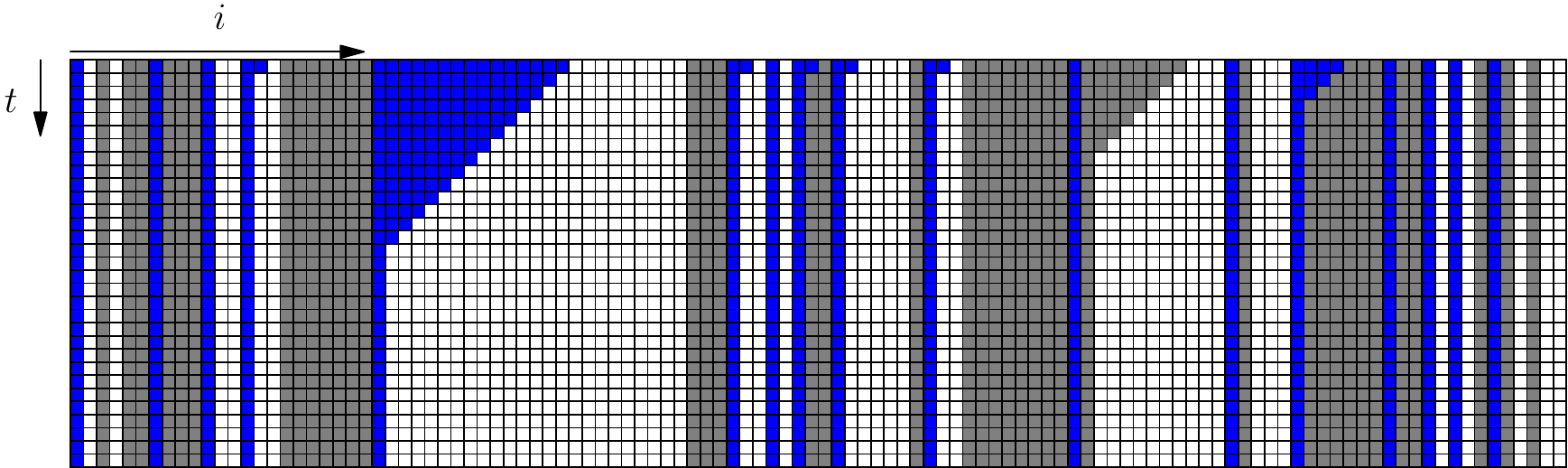}
 \end{center}
\caption{Sample spatio-temporal pattern generated by 3-state rule 140. White, lighter gray and
darker gray cells (blue in color version) correspond, respectively, to 0, 1 and~2.} \label{samplepatern}
\end{figure}

We studied dynamics of $f_g$ for a number of elementary rules $g$. One of the most interesting of them
is the case of $g$ being elementary CA rule with Wolfram number 140, defined as
\begin{equation} \label{rule140}
 g(x_1,x_2,x_3) = x_2-x_1 x_2+x_1 x_2 x_3,
\end{equation}
where $x_1,x_2,x_3 \in \{0,1\}$. As we will see, it actually exhibits degenerate hyperbolicity. In what follows,
we will refer to $f_g$ with $g$ given by eq. (\ref{rule140})  as ``rule 140''. We will also drop the index $g$ and
refer to $f_g$ simply as $f$. An example of a spatio-temporal pattern produced by this rule is shown in
Figure~\ref{samplepatern}.

%

Let us first introduce the notion of density polynomials.
Let ${\mathcal{A}}=\{0,1,2\}$.
A finite sequence of elements of ${\mathcal{A}}$, $\mathbf{b}=b_1b_2\ldots, b_{n}$, will be called a \emph{block} 
 (or \emph{word})  of length $n$. The set of all blocks of elements of ${\mathcal{A}}$ of all possible lengths will be denoted by ${\mathcal{A}}^{\star}$.

A {\em block evolution operator} corresponding to $f$ is a mapping
 $\f:\B \mapsto \B$ defined as follows. Let  $\mathbf{a}=a_1a_2 \ldots a_{n}\in \G^n$
where $n \geq 3$. Then $\f(\mathbf{a})$ is a block of length $n-2$ defined as
\begin{equation}
\f(\mathbf{a}) = f(a_1,a_{2},a_{3})
f(a_2,a_{3},a_{4})\ldots
f(a_{n-2},a_{n-1},a_{n}).
\end{equation}
If  $\mathbf{f}(\mathbf{b})=\mathbf{a}$, than we will say that $\mathbf{b}$ is a preimage of
$\mathbf{a}$, and write $\mathbf{b} \in \mathbf{f}^{-1}(\mathbf{a})$.
Similarly, if $\mathbf{f}^n(\mathbf{b})=\mathbf{a}$, than we will say that $\mathbf{b}$ is an
\emph{$n$-step preimage} of
$\mathbf{a}$, and write $\mathbf{b} \in \mathbf{f}^{-n}(\mathbf{a})$.

Let the \emph{density polynomial} associated with a  string $\mathbf{b}=b_1b_2\ldots b_n$ be defined 
as 
\begin{equation}
 \Psi_{\mathbf{b}}(p,q,r)=p^{\#_0 (\mathbf{b})} q^{\#_1 (\mathbf{b})} r^{\#_2 (\mathbf{b})},
\end{equation}
where $\#_i (\mathbf{b})$ is the number of occurrences of symbol  $i$ in $\mathbf{b}$.
If $A$ is a set of strings, we define density polynomial associated with $A$ as
\begin{equation}
 \Psi_{A}(p,q,r)=\sum_{\mathbf{a} \in A} \Psi_{\mathbf{a}}(p,q,r).
\end{equation}

One can easily show (in a manner similar as done in \cite{paper52}) that if one starts with a bi-infinite string of symbols drawn from 
Bernoulli distribution where probabilities of $0,1$ and $2$ are, respectively, $p,q$ and $r$, then
the proportion of sites in state $k$  after $n$ iterations of rule $f$
 is given by $\Psi_{\f^{-n}(k)}(p,q,r)$.
This quantity will be called \emph{density} of symbols $k$ after $n$ iterations of $f$. 

For 3-state rule 140 defined by eqs. (\ref{defoflift}) and (\ref{rule140}), we generated sets of $n$-step preimages of 0, 1 and 2 for $n$ varying from 1 to 7. 
Using AT\&T FSM Library \cite{fsmlib}, we constructed minimal finite state machines (FSM) generating
these sets, and we found that these FSM exhibit regularities which can be exploited
to produce general expressions for  $\Psi_{\f^{-n}(k)}(p,q,r)$. Results are described below.
Formal proofs are omitted for lack of space, but they are available upon request and will be published elsewhere.
\section{Structure of preimages of 1}
\begin{figure}[t]
(a)\includegraphics[scale=0.5]{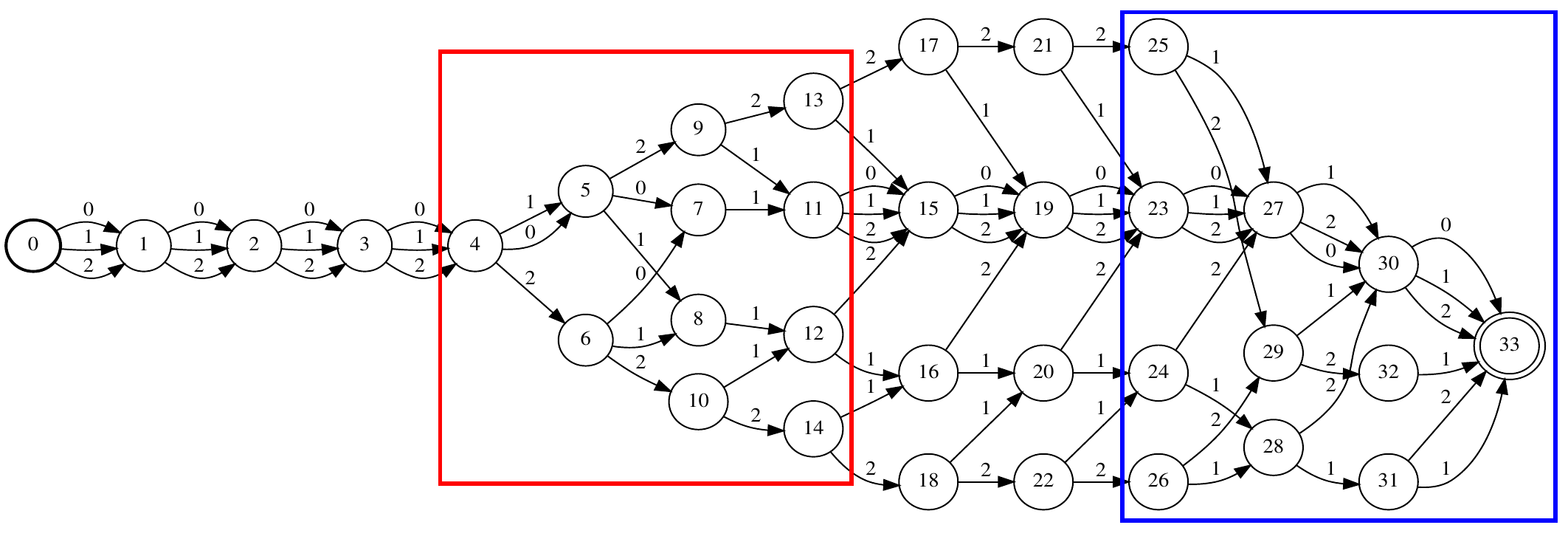} \\[3em]
(b)\includegraphics[scale=0.55]{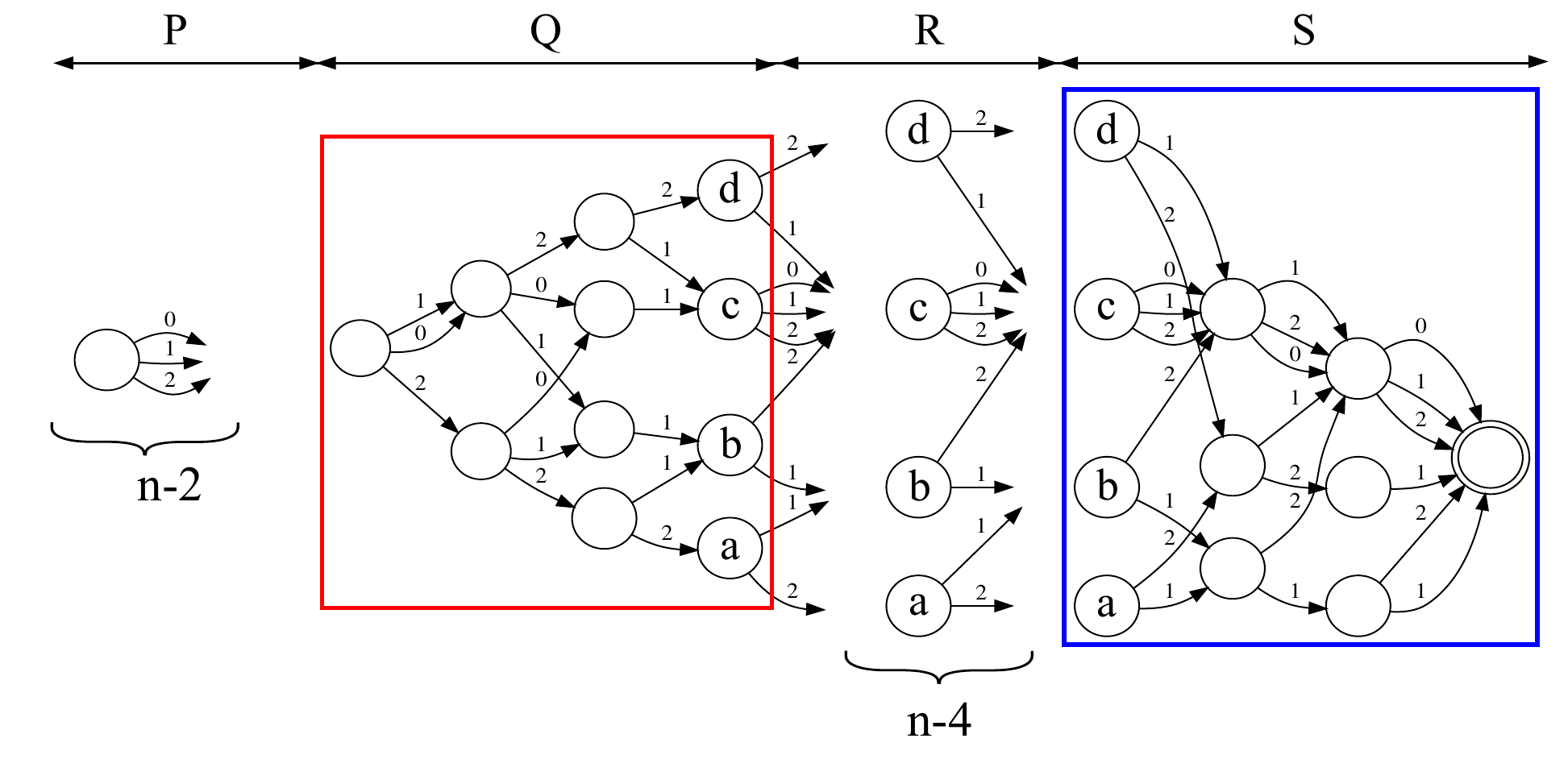} 
\caption{Finite state machine for for $6$-step (a) and (b)  $n$-step preimages of 1 under the rule 140.}\label{generalpreim}
\end{figure}

The set of $n$-step perimages of 1 can be described by a finite state machine (FSM) schematically
shown in Figure~\ref{generalpreim}. The FSM has four parts, denoted by P,Q,R  and S. Parts
Q and S are always the same, while parts P and R consists of repeated graph fragments, where the number
of repetitions is, respectively, $n-2$ and $n-4$.

Let us consider part Q first. If we start from the leftmost node of P and wish to end at the node labeled
by $a$, the only path with such property is $222$, corresponding to density polynomial
$r^3$. Similarly, if we want to end at $b$, the possible paths are $\star11$ and $221$, yielding density
polynomial $(p+q+r)q^2 + r^2q$. For the node $c$, the  possible paths are
$021, 121, 001, 101$ and $201$, so the density polynomial is $2pqr+q^2r+p^2q+q^2p$. Finally, for node $d$,
the paths are $022$ and $122$, so that the polynomial is $pr^2+qr^2$.
Let us, therefore, define a vector with entries corresponding to 
density polynomials of paths ending at $a$, $b$, $c$, and $d$, 
\begin{equation}
 Q=\left[ \begin {array}{c} {r}^{3}\\ \noalign{\medskip} \left( p+q+r
 \right) {q}^{2}+{r}^{2}q\\ \noalign{\medskip}2\,pqr+{q}^{2}r+{p}^{2}q
+p{q}^{2}\\ \noalign{\medskip}p{r}^{2}+{r}^{2}q\end {array} \right]. 
\end{equation}
In a very similar fashion, we can construct a vector holding density polynomials for paths
starting at nodes of segment S labeled, respectively,  $a$, $b$, $c$, and $d$, and ending at the
rightmost node,
\begin{equation}
 S=\left[ \begin {array}{c} {q}^{3}+{q}^{2}r+2\, \left( p+q+r \right) qr
+{r}^{2}q\\ \noalign{\medskip}{q}^{3}+{q}^{2}r+ \left( p+q+r \right) q
r+r \left( p+q+r \right) ^{2}\\ \noalign{\medskip} \left( p+q+r
 \right) ^{3}\\ \noalign{\medskip}{r}^{2}q+ \left( p+q+r \right) qr+
 \left( p+q+r \right) ^{2}q\end {array} \right]. 
\end{equation}
It is easy to verify that the above components of S correspond to paths starting from $a$ ($111,112,12\star,
21\star,221$), from $b$ ($111, 112, 12\star, 2\star\star$), from $c$ ($\star\star\star$) and
from $d$ ($221, 21\star, 1\star\star$).

Let us now analyze segment R of the FSM shown in Figure~\ref{generalpreim}. First let us suppose that
there is no repeated part in segment $R$, as it would be for the case of $n=4$, when the number of repetitions is 
$n-4=0$. Let us
construct a matrix $R$ such that $R_{i,j}$  represents the density polynomial
of all paths starting from node $j$ of segment Q and ending in node $i$ of segment S, where
$i,j\in \{a,b,c,d\}$. This matrix has the form
\begin{equation}
R= \left[ \begin {array}{cccc} r&0&0&0\\ \noalign{\medskip}q&q&0&0
\\ \noalign{\medskip}0&r&p+q+r&q\\ \noalign{\medskip}0&0&0&r
\end {array} \right].
\end{equation}
As we can see, the only non-zero entries are diagonal ones and $R_{b,a}=q$, $R_{c,b}=r$, $R_{c,d
}=q$. This is because there are only three ways to finish at a different node that we started,
namely if we start from $a$ and finish at $b$ (generating symbol $1$ along the way),
if we  we start from $b$ and finish at $c$ (generating 2), or if  we start from $d$ and finish at $c$ (generating 1).  

Suppose now that the repeated fragment in segment R is repeated $n-4$ times. It is not hard to see that
density polynomials for all paths from node $j$ of segment Q to node $i$ of segment S
will be represented by entries of matrix $R^{n-3}$. Furthermore, all paths from the beginning of
segment $Q$ to the end of segment S will be represented by the density polynomial
given by $S^{T} R^{n-3} Q$, where $T$ denotes transposition (row vector). Since the segment  P
is represented by $(p+q+r)^{n-2}$, the final expression for the density polynomial of 
$n$-step preimages of 1 is
\begin{equation}
 \Psi_{\f^{-n}(1)}(p,q,r)=(p+q+r)^{n-2} S^{T} R^{n-3} Q.
\end{equation}
In order to obtain more explicit expression for $\Psi(p,q,r)$, we will need to compute $R^{n-3}$.
When $r\neq q$, $R$ is invertible, and one can diagonalize it, 
\begin{equation}
 R=L  \left[ \begin {array}{cccc} q&0&0&0\\ \noalign{\medskip}0&r&0&0
\\ \noalign{\medskip}0&0&p+q+r&0\\ \noalign{\medskip}0&0&0&r
\end {array} \right] 
L^{-1},
\end{equation}
where
\begin{equation}
 L=\left[ \begin {array}{cccc} 0&{\frac {q}{r}}&0&{\frac {q-r}{r}}
\\ \noalign{\medskip}{\frac {q}{q-r}}&-{\frac {{q}^{2}}{r \left( q-r
 \right) }}&0&-{\frac {q}{r}}\\ \noalign{\medskip}-{\frac {qr}{
 \left( p+r \right)  \left( q-r \right) }}&{\frac {qr}{ \left( q-r
 \right)  \left( p+q \right) }}&{\frac {qr}{{p}^{2}+pq+pr+qr}}&0
\\ \noalign{\medskip}0&1&0&1\end {array} \right]. 
\end{equation}
This yields, after simplification,
\begin{multline} \label{normalpreim1}
\Psi_{\f^{-n}(1)}(p,q,r)= {\frac {  p {q}^{2} \left( -pr+pq+{q}^{2} \right)
\left( q \lambda \right)^{n} }{ \lambda^{2} \left( p+r \right)  \left( q
-r \right) }}\\
+{\frac {q r \left( -{p}^{2}r+{p}^{2}q+p{q}^{2}-2\,pqr+{r}^
{3}-{q}^{2}r \right)   \left( r \lambda \right)^{n}}{ 
\lambda^{2} \left( p+q \right)  \left( q-r \right) }}\\
+{\frac {q
 \left( {p}^{3}+{p}^{2}q+2\,{p}^{2}r+p{r}^{2}+3\,pqr+{r}^{3}+{r}^{2}q+
{q}^{2}r \right)  \lambda^{2n} }
 { \lambda  \left( p+r \right)  \left( p+q
 \right) }},
\end{multline}
where we used $\lambda=p+q+r$.

When $r=q$, matrix $R$ becomes singular. In can be written in Jordan form as
\begin{equation}
 R=L (M+N) L^{-1},
\end{equation}
where 
\begin{equation}
 L= \left[ \begin {array}{cccc} 0&0&1&0\\ \noalign{\medskip}0&q&-1&-1
\\ \noalign{\medskip}{\frac {{q}^{2}}{ (p+q)^2 } }&-{\frac 
{{q}^{2}}{p+q}}&-{\frac {{q}^{2}}{(p+q)^2}}&0
\\ \noalign{\medskip}0&0&1&1\end {array} \right], \,\,\,
 M= \left[ \begin {array}{cccc} p+2\,q&0&0&0\\ \noalign{\medskip}0&q&0&0
\\ \noalign{\medskip}0&0&q&0\\ \noalign{\medskip}0&0&0&q\end {array}
 \right], \,\,\, 
N= \left[ \begin {array}{cccc} 0&0&0&0\\ \noalign{\medskip}0&0&1&0
\\ \noalign{\medskip}0&0&0&0\\ \noalign{\medskip}0&0&0&0\end {array}
 \right].
\end{equation}
Matrices $M$ and $N$ commute, and matrix $N$ is nilpotent, $N^2=0$. Because of this,
for any integer $k$, 
\begin{equation}
 (M+N)^k=M^k + kNM^{k-1} =  \left[ \begin {array}{cccc}  \left( p+2\,q \right) ^{k}&0&0&0
\\ \noalign{\medskip}0&{q}^{k}&0&0\\ \noalign{\medskip}0&0&{q}^{k}&0
\\ \noalign{\medskip}0&0&0&{q}^{k}\end {array} \right] 
+
 \left[ \begin {array}{cccc} 0&0&0&0\\ \noalign{\medskip}0&0&k{q}^{k-1
}&0\\ \noalign{\medskip}0&0&0&0\\ \noalign{\medskip}0&0&0&0
\end {array} \right], 
\end{equation}
and finally
\begin{multline} \label{degeneratepreim1}
 \Psi_{\f^{-n}(1)}(p,q,q)=(p+q+q)^{n-2} S^{T} R^{n-3} Q=
(p+2q)^{n-2} S^{T} L (M+N)^{n-3} L^{-1} Q
\\= (p+2q)^{n-2} S^{T} L
 \left[ \begin {array}{cccc}  \left( p+2\,q \right) ^{n-2}&0&0&0
\\ \noalign{\medskip}0&{q}^{n-2}&{q}^{n-3} \left( n-2 \right) &0
\\ \noalign{\medskip}0&0&{q}^{n-2}&0\\ \noalign{\medskip}0&0&0&{q}^{n-
2}\end {array} \right] 
 L^{-1} Q
\end{multline}
After simplification this yields
\begin{multline} \label{P1degenerate}
\Psi_{\f^{-n}(1)}(p,q,q)= {\frac {p{q}^{3} \left( n+1 \right)  \left( q \lambda \right) ^{n}
}{ \lambda^{2} \left( q+p \right) }}
+{\frac {{q}^{2}
 \left( 2\,{p}^{3}+4\,{p}^{2}q+p{q}^{2}-2\,{q}^{3} \right)  \left( q \lambda \right) ^{n}}{ \left( q+p \right) ^{2}  \lambda^{2}}}\\
+{\frac { \left( {p}^{3}+3\,{p}^{2}q+4\,p{q}^{2}+3\,{q}
^{3} \right) q \lambda^{2n}}{ 
 \lambda   \left( q+p \right) ^{2}}},
\end{multline}
where, as before, $\lambda=p+q+r=p+2q$.

We shall add here that  even though eq. (\ref{normalpreim1}) and (\ref{P1degenerate}) were derived assuming $n\geq 4$,
 they happen to be correct for $n=1,2$ and $3$ as well. Let us also remark that
when one substitutes $p=1,q=1$ and $r=1$, then $\Psi_{\f^{-n}(1)}(1,1,1)$ counts the number of preimages of~1. This yields
a sequence exhibiting linear-exponential growth,
\begin{equation}
 \Psi_{\f^{-n}(1)}(1,1,1)=\left(\frac{n}{18}+ \frac{7}{36} \right) 3^n + \frac{11}{12}9^n.
\end{equation}
\section{Structure of preimages of 0 and 2}
\begin{figure}[t]
a) \includegraphics[scale=0.55]{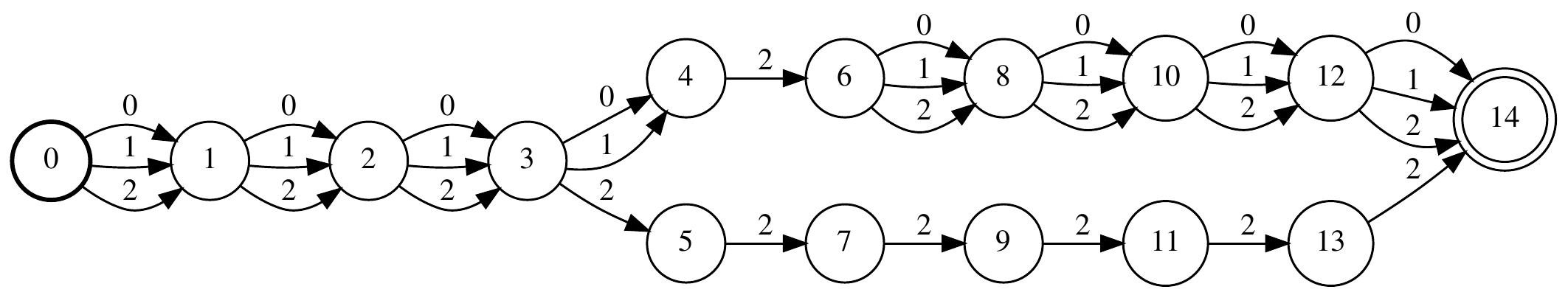}\\[2em]
b) \includegraphics[scale=0.85]{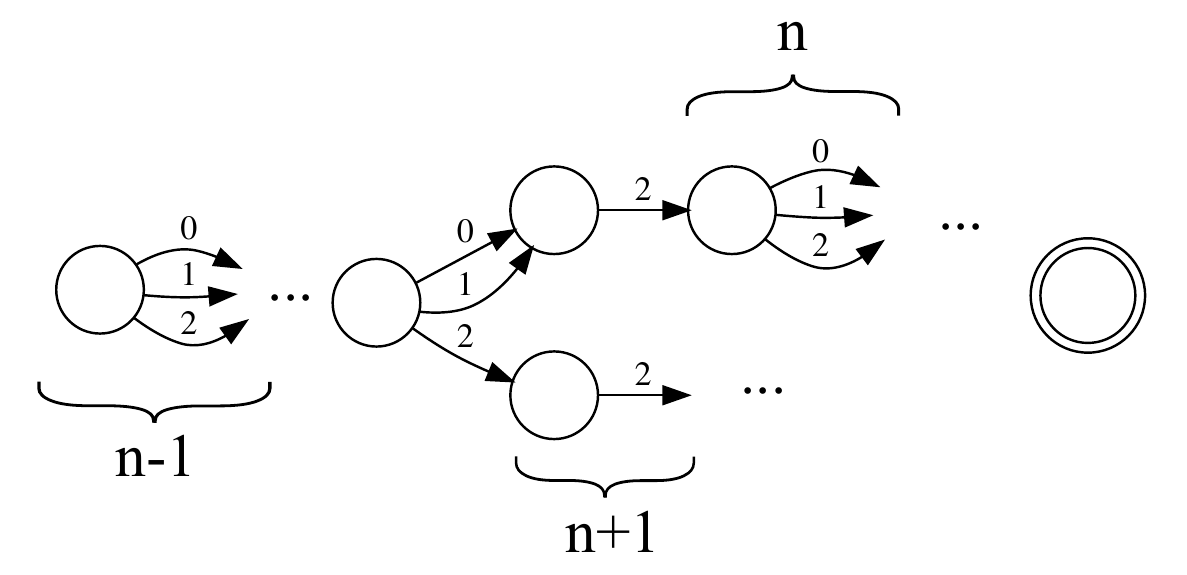} 
\caption{Finite state machines for $4$-step (a) and $n$-step (b) preimages of 2 under the rule 140.} \label{preimagesof2}
\end{figure}
For preimages of 0, FSM generating preimage sets are quite similar as for preimages of 1, thus we will omit details. Similar analysis
as in the previous section yields, for $r \neq q$,
\begin{multline} \label{normalpreim0}
\Psi_{\f^{-n}(0)}(p,q,r)= {\frac { \left( -pr+pq+{q}^{2} \right) p{q}^{2} \left(q \lambda
 \right) ^{n}}{ \lambda^{2} \left( p+r \right)  \left( 
r-q \right) }}
+{
\frac {p r \left( -{r}^{2}p+{q}^{2}p+{q}^{3}-{r}^{3}-q{r}^{2} \right) 
 \left( r \lambda \right) ^{n}}{ \lambda^{2} \left( 
p+q \right)  \left( r-q \right) }}
\\
+{\frac { \left( {p}^{3}+2\,{p}^{2}q+2\,{p}^{2}r+2\,{r}^
{2}p+3\,qpr+2\,{q}^{2}p+{r}^{3}+2\,q{r}^{2}+{q}^{3}+{q}^{2}r \right) p
 \lambda^{2n}}{
 \left( p+q \right)  \left( p+r \right)  \lambda }},
\end{multline}
 and for $r=q$,
\begin{multline} \label{P0degenerate}
\Psi_{\f^{-n}(0)}(p,q,q) ={\frac { \left( {p}^{3}+4\,{p}^{2}q+7\,{q}^{2}p+5\,{q}^{3} \right) p
 \lambda^{2n}}{ \lambda 
 \left( p+q \right) ^{2}}}
-{\frac {p{q}^{3} \left( n+1 \right) 
 \left( q \lambda \right) ^{n}}{  \lambda^{2}
 \left( p+q \right) }}
\\
-{\frac {{q}^{2}p \left( 3\,{p}^{2}+8\,pq+6\,{q}
^{2} \right)  \left( q \lambda \right) ^{n}}{ \left( p+q \right) ^
{2} \lambda  ^{2}}}.
\end{multline}
Again, when $p=q=r=1$, the density polynomial counts preimages of 0, and we obtain
 \begin{equation}
  \Psi_{\f^{-n}(0)}(1,1,1)=\frac{17}{12} 9^n -\left(\frac{n}{18}+\frac{19}{36}\right) 3^n.
 \end{equation}
This sequence, similarly as the number of preimages of 1, exhibits linear-exponential growth.

For preimages of 2, preimage sets have much simpler structure, shown in Figure~\ref{preimagesof2}. Density polynomials 
for them are given by
\begin{equation} \label{P2}
\Psi_{\f^{-n}(2)}(p,q,r)=  \left( q+p \right) r \lambda^{2n-1}
+{\frac {{r}^{2} }{\lambda}} \left( r \lambda \right) ^{n}.
\end{equation}
The number of preimages, obtained by taking $p=q=r=1$, is in this case 
 \begin{equation}
 \Psi_{\f^{-n}(2)}(1,1,1)=3^{n-1}+\frac{2}{3} 9^n,
 \end{equation}
thus no linearity is present.
\section{Conclusions}
As mentioned in the introduction, density polynomials $\Psi_{\f^{-n}(k)}(p,q,r)$ represent probability of occurence
of $k$ after $n$ iterations starting from a Bernoulli distribution with probabilities of 0, 1 and 2 equal to, respectively, 
$p$, $q$, and $r$, where $p+q+r=1$. If we start with a symmetric Bernoulli distribution
where $r=q$, the probability of occurence of 1 after $n$ steps, to be denoted by $P_n(1)$, will be given by eq. (\ref{P1degenerate}) in which we substitute $r=q$ and $q=(1-p)/2$. This yields, after simplification,
\begin{equation}
P_n(1)=P_{\infty}(1)-\frac{\left( p-1 \right) ^{2}}{4 \left( 1+p
 \right) ^{2}} \, \left( {p}^{3}n-{p}^{3}-pn-5\,p-3\,{p}^{2}+1 \right) 
  \left(\frac{1-p}{2} \right) ^{n},
\end{equation}
where
\begin{equation}
 P_{\infty}(1) = {\frac { \left( 1-p \right)  \left( {p}^{3}+5\,{p}^{2}-p+3
 \right) }{4 \left( 1+p \right) ^{2}}}.
\end{equation}
It is clear that for $0<p<1$, $P_n(1)$ tends to  $P_{\infty}(1)$ as $n \to \infty$, and that the convergence is linear-exponential 
in $n$. Such ``degenerate'' convergence takes place for  probability of occurence of 0 as well, as seen in
eq. (\ref{P0degenerate}). When $r \neq q $ in the initial Bernoulli distribution, the convergence
is purely exponential, as in eq. (\ref{normalpreim1}) and (\ref{normalpreim0}). Probability of occurence of 2 is also 
always exponential, and degeneracy is not possible in this case.

The example of 3-state rule  presented here is an interesting instance of a phenomenon similar to degenerate 
hyperbolicity in finite-dimensional dynamical systems. It is hoped that it stimulates further research
on hyperbolicity in CA. The method of constructing density polynomials by using finite state machines appears to
be quite fruitful, and it should be applicable to many other cellular automata rules.


\begin{thebibliography}{1}
\providecommand{\url}[1]{\texttt{#1}}
\providecommand{\urlprefix}{URL }

\bibitem{fsmlib}
{AT\&T Finite-State Machine Library}, version 4.0,
  \url{http://www3.cs.stonybrook.edu/~algorith/implement/fsm/implement.shtml}

\bibitem{probcabookchapter}
Fuk\'s, H.: An example of computation of the density of ones in probabilistic
  cellular automata by direct recursion (2015), submitted for publication

\bibitem{paper52}
Fuk\'s, H., Soto, J.M.G.: Exponential convergence to equilibrium in cellular
  automata asymptotically emulating identity. Complex Systems  23,  1--26
  (2014)

\end{thebibliography}
\end{document}